\documentclass[RNAAS]{aastex62}

\begin{document}

\title{White Dwarf Collisions, a promising scenario to account for meteoritic anomalies}  

\correspondingauthor{Jordi Isern}
\email{isern@ice.cat, eduardo.bravo@upc.edu}

\author{Jordi Isern}
\affiliation{ICE,CSIC and IEEC}

\author{Eduardo Bravo}
\affiliation{Department of Physics (UPC)}

\keywords{stars: white dwarfs - nucleosynthesis; meteorites}

 \section{}
It is commonly accepted that collisions between white dwarfs are rare events that only occur in the dense interior of globular clusters or in the dense outskirts around the central galactic black holes, and are therefore disregarded as an important source of Type Ia supernovae (SNIa). This perception recently changed when it was realized that the majority of stars are not single but forming larger systems.

Triple systems are relatively common and those with an hierarchical structure are stable. However, the presence of a third star around the binary can periodically modify the eccentricity of the inner orbit and the inclination of the outer orbit (the Kozai-Lidov cycles) increasing the number of encounters and collisions that are strong enough to produce SNIa events. In fact it has been claimed that these systems, the hierarchical triples, could account for an important number of SNIa \citep{katz12}. See \citet{toon18} for a different point of view.

Although the majority of these encounters will not result in a SNIa event, many of them will produce mass ejections. The ejections consist of pristine, unprocessed, WD material and products resulting from the nuclear reactions triggered by the collision. The abundance and nature of these contaminants, made of intermediate mass elements (IME) to $^{56}$Ni, depend on the strength of the collision and on the nature of WDs \citep{rask09}.
Under the appropriate circumstances, this material can become part of a protostar nebula, including the pre-solar one, in the form of stardust leading to the existence of chemical anomalies in meteorites. A collision with a neutron star or a black hole could produce similar effects. 

To see the potential of such scenarios consider the so called Ne-E anomaly found in some primitive meteorites like Orgueil and Murchison \citep{blac69,amar09}. This anomaly is based on the existence of graphite grains of pre-solar origin that contain an excess of $^{22}$Ne as compared with solar values. In this case, the challenge is how to produce such excesses within the oxygen-poor environtment that is necessary for the existence of graphite. 

Figure 1 displays the chemical structure of a typical carbon oxygen 0.64 M$\odot$ WD (dashed lines) at the beginning of the cooling process, assuming to simplify that is only made of C/O/Ne and that the abundance of $^{22}$Ne is nearly equal to the sum of the initial C+N+O content. During this process, the chemical profile gradually changes as a consequence of the gravitational settling of heavier chemical species, mainly neutron rich isotopes, and sedimentation induced by the changes of solubility upon crystallization \citep{iser97}. Figure 1 shows (continuous lines) the chemical profiles predicted by the \citet{segr96} phase diagram when the crystallization of the star is complete and the influence of gravitational diffusion is neglected. The abundance of oxygen increases in the central regions while carbon is expelled to the outer layers, to the point that it becomes more abundant than oxygen, while the abundance of neon is unaffected in these regions. When carbon exceeds $\approx 0.75$, oxygen is expelled from the mixture and C/Ne displays azeotropic behavior leading to the formation of a thin layer made of carbon and neon. See \citet{segr96} for a detailed description of the process. 

\begin{figure}[h!]
\begin{center}
\includegraphics[scale=0.8,angle=0]{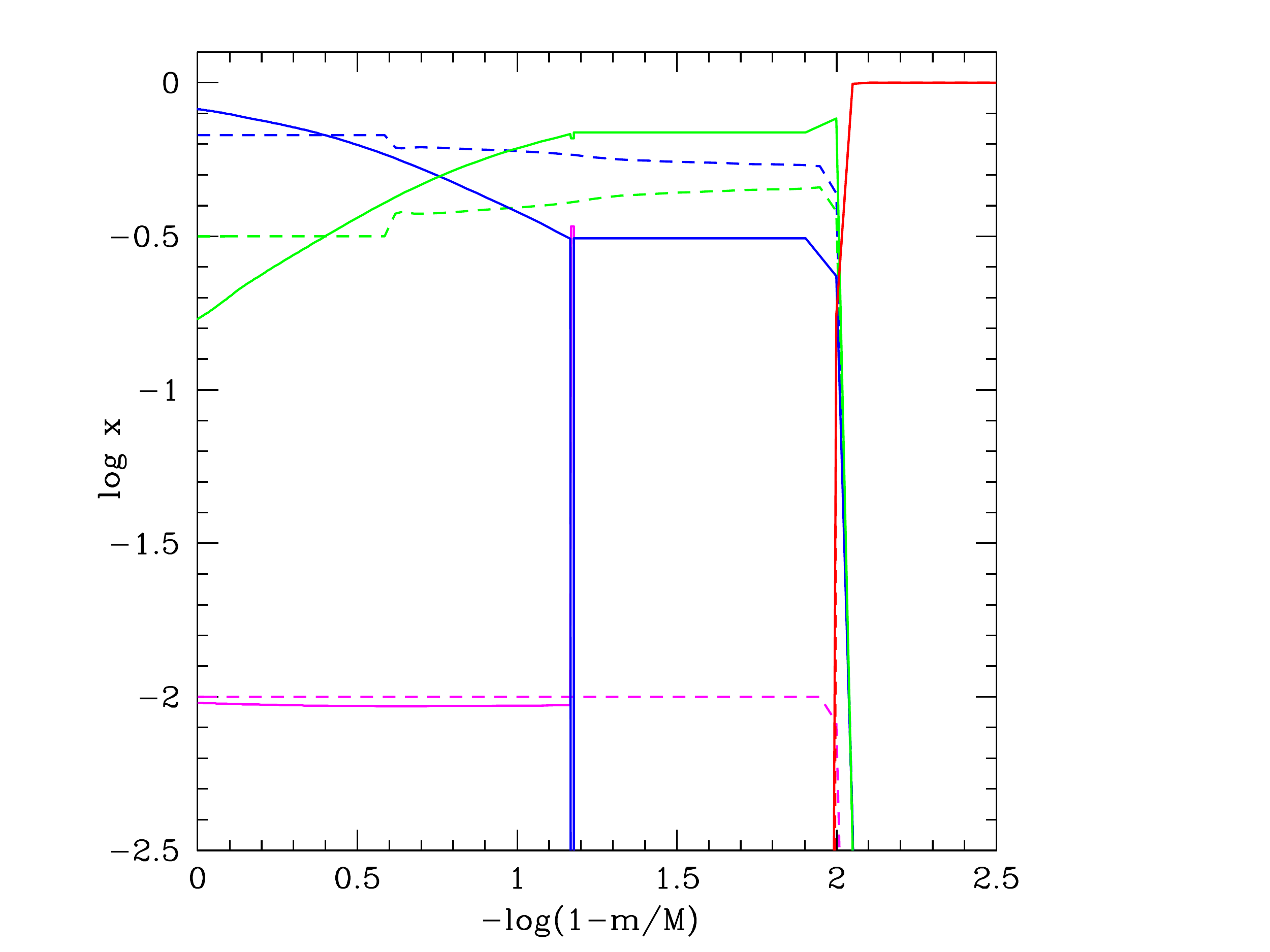}
\caption{Chemical profile of a typical white dwarf where m and M are the mass coordinate and total mass respectively. Lines represent the mass fraction of $^4$He (red), $^{12}$C (green), $^{16}$O (blue), and $^{22}$Ne (magenta). Dashed lines display the initial chemical composition and solid ones what is obtained after complete crystallization using the C/O/Ne phase diagram of \citet{segr96}. The mass of the C/Ne layer is $1.37 \times 10^{-3}$ M$_\odot$ and the $^{22}$Ne abundance is the azeotropic one, $x_{22} =0.34$  in fraction number.\label{fig:1}}

\end{center}
\end{figure}

As an example, the collision of two C/O-WD of 0.64 M$_\odot$ each, with a relative velocity of 6000 kms$^{-1}$ and impact parameter equal to the radius of the WD ejects 0.06 M$_\odot$ containing $2.2\times 10^{-4}$ M$\odot$ of the C/Ne material with the azeotropic composition, and 0.011 and $3.2 \times 10^{-4}$ M$_\odot$ of freshly synthesized $^{28}$Si and $^{56}$Ni respectively (Isern et al. in preparation). In order to evaluate the impact of this scenario it is necessary to obtain the detailed nucleosynthesis for different masses, initial chemical composition and collision parameters, as well as to elucidate the behavior of other impurities like  $^{13}$C, $^{18}$O, $^{20}$Ne, $^{28}$Si, and s-elements, the  formation of dust during such events, and the limits imposed by the chemical evolution of the Galaxy to the WD collision frequency.

\acknowledgments

This work  has been supported  by MINECO grants  ESP2013-47637-P (JI) and ESP2015-63588-P (EB), by the  European Union FEDER funds, and by the grant 2014SGR1458 (JI)  of  the Generalitat de Catalunya.

\end{document}